\documentclass[twoside]{articlek}

\usepackage[normal]{caption}

\renewcommand{\refname}{REFERENCES}

\def\BibTeX{{\rm B\kern-.05em{\sc i\kern-.025em b}\kern-.08em
            T\kern-.1667em\lower.7ex\hbox{E}\kern-.125emX}}

\usepackage{graphicx}

\begin{document}

\begin{flushleft}
{\large\bf 
SYMMETRY ENERGY: FROM NUCLEAR MATTER TO FINITE NUCLEI}
\vspace*{25pt} 

{\bf V. M. Kolomietz and A. I. Sanzhur}\\
\vspace{5pt}
{Institute for Nuclear Research, NAS of Ukraine, Prospekt Nauky 47, 03680 Kyiv, Ukraine}\\

\end{flushleft}

%---------------------------------------------------------------------------
\vspace{5pt}
\begin{abstract}\noindent
We suggest a particular procedure of derivation of the beta-stability line
and isotopic symmetry energy. The behavior of the symmetry energy coefficient
$b(A,N-Z)$ is analyzed. We redefine the surface tension coefficient and the
surface symmetry energy for an asymmetric nuclear Fermi-liquid drop with a finite
diffuse layer. Following Gibbs-Tolman concept, we introduce the equimolar
radius at which the surface tension is applied. The relation of the nuclear
macroscopic characteristics like surface and symmetry energies, Tolman length,
etc. to the bulk properties of nuclear matter is considered.
The surface-to-volume symmetry energy ratio for several Skyrme-force
parametrizations is obtained.
\end{abstract}
\vspace{5pt}
%---------------------------------------------------------------------------

\section{INTRODUCTION}
Many static and dynamic features of nuclei are sensitive to the symmetry
energy and the isospin degrees of freedom. The basic characteristics of
isovector giant and isobar analog resonances \cite{bomo2}, the isoscaling in
nuclear fission and multifragmentation
\cite{tsfr01,frid04,veso04,lili04,sosh03,veso04a,gebr04,mawa05}
depend significantly on the isospin effects. The symmetry energy is also a key
element for the derivation of the nuclear stability valley. The nuclear
$\beta$-stability is determined by the balance of the isotopic symmetry,
$E_{\mathrm{sym}}$, and the Coulomb, $E_{C}$, energies. However the extraction
of both $E_{\mathrm{sym}}$ and $E_{C}$ from the nuclear binding energy is
not a simple problem because of their complicate dependency on the mass
number $A=N+Z$ in finite nuclei with $N$ neutrons and $Z$ protons \cite{auwa03}.
The standard procedure of extraction of the symmetry energy from a
fit of mass formula to the experimental binding energies \cite{jado03} is
not free from ambiguities and does not allow one to separate the symmetry
energy into the volume and surface contributions
directly.
In the present work, to study the structure of the $\beta$-stability line
and both $E_{\mathrm{sym}}$ and $E_{C}$ energies we use a particular
procedure which is based on the dependence of the isospin shift of
neutron-proton chemical potentials $\Delta\lambda (X)=\lambda_{n}-\lambda_{p}$
on the asymmetry parameter $X=(N-Z)/(N+Z)$ for nuclei beyond the
$\beta$-stability line. This procedure allows us to represent the results
for the $A$-dependence of the $\beta$-stability line and both energies
$E_{\mathrm{sym}}$ and $E_{C}$ in a transparent way, which can be easily used
for the extraction of the smooth volume and surface contributions as well as
their shell structure. Note also that our procedure of extraction of all
values $E_{C}(A)$ and $E_{\mathrm{sym}}$ is partly model independent, that is,
the theoretical models for calculations of the nuclear binding energy as well
as the nucleon distributions are not involved. We only assume the commonly
used parabolic dependence of the symmetry energy on the asymmetry parameter $X$.
Due to the charge invariance of the nuclear forces this assumption is well
justified for small values of $X$. A similar approach based on the isobaric
multiplet mass equation \cite{orma97} was used in Ref.~\cite{liwa11}
to study the Coulomb parameter within the modern nuclear mass model WS3.

The nucleus is a two component, charged system with a finite diffuse layer.
This fact specifies a number of various peculiarities of the nuclear surface
and symmetry energies: dependency on the density profile function, non-zero
contribution to the surface symmetry energy, connection to the nuclear
incompressibility, etc. The additional refinements appear due to the quantum
effects arising from the smallness of nucleus. In particular, the curved
interface creates the curvature correction to the surface energy
$E_{\mathcal{S}}$ and the surface part of symmetry energy $E_{\mathrm{sym}}$ of
order $A^{1/3}$ and can play the appreciable role in small nuclei as well as
in neck region of fissionable nuclei.
The presence of the finite diffuse layer in nuclei creates the problem of
the correct definition of the radius and the surface of tension for a small
drop with a diffuse interface. Two different radii have to be introduced in
this case \cite{gibbs, tolm49}: the equimolar radius $R_{e}$, which gives the
actual size of the corresponding sharp-surface droplet, and the radius of
tension $R_{s}$, which derives, in particular, the capillary pressure.
Bellow we will address this problem to the case of two-component nuclear
drop. In general, the presence of the curved interface affects both the bulk
and the surface properties. The curvature correction is usually negligible
in heavy nuclei. However, this correction can be important in some nuclear
processes. For example the yield of fragments at the nuclear
multifragmentation or the probability of clusterization of nuclei from the
freeze-out volume in heavy ion collisions \cite{kosa12}. In both above
mentioned processes, small nuclei necessarily occur and the exponential
dependence of the yield on the surface tension \cite{lali58} should cause a
sensitivity of both processes to the curvature correction. Moreover the
dependency of the curvature interface effects on the isotopic asymmetry of
small fragments can significantly enhance (or suppress) the yields of
neutron rich isotopes.
In this paper we analyze of the interface effects in an asymmetric
nuclear Fermi-liquid drop with a finite diffuse layer. We follow the
ideology of the extended Thomas-Fermi approximation (ETFA) with effective
Skyrme-like forces combining the ETFA and the direct variational method with
respect to the nucleon densities, see Ref.~\cite{kosa08}.
The surface and
symmetry energies were widely studied earlier taking into consideration also the
finite surface thickness and the curvature corrections
\cite{mysw69, mysw85, cent98, cero10, roce11, lape85}. Note also the applications
of the ETFA with the Skyrme-type interactions to the studies of the nuclear bulk,
surface and symmetry properties, see e.g. Refs. \cite{brgu85,cevi93,trkr86,kopr85}.
In order to formulate proper definition for
the drop radius, we use the concept of the dividing surface, originally
introduced by Gibbs \cite{gibbs}. Following the Gibbs method, which is
applied to the case of two component system, we introduce the superficial
(surface) density as the difference (per unit area of dividing surface)
between actual number of particles $A$ and the number of bulk,
$A_{\mathcal{V}}$, and  neutron excess, $A_{-,\mathcal{V}}$, particles which a
drop would contain if the particle densities were uniform.

\section{STRUCTURE OF $\beta$-STABILITY LINE AND SYMMETRY ENERGY}
Considering the asymmetric nuclei with a small asymmetry parameter
$X=(N-Z)/A\ll 1$, the total energy per nucleon $E/A$ can be represented in
the following form of $A,X$-expansion, 
\begin{equation}
E/A=e_{0}(A)+b(A)X^{2}+E_{C}(X)/A\ ,
\label{E/A}
\end{equation}
where $e_{0}(A)$ includes both the bulk and the surface energies,
$b(A)$ is the symmetry energy coefficient, $E_{C}(X)$ is the
total Coulomb energy 
\begin{equation}
E_{C}(X)/A=e_{C}(A)(1-X)^{2}\ .
\label{eC}
\end{equation}
Using the derivation of the chemical potential $\lambda_{q}$ ($q=n$ for a
neutron and $q=p$ for a proton)
\begin{equation}
\lambda_{n}=\left( \frac{\partial E}{\partial N}\right)_{Z}\ ,\ \ \
\lambda_{p}=\left( \frac{\partial E}{\partial Z}\right)_{N}\ ,
\label{dlamb}
\end{equation}
one can write the condition of nuclear $\beta$-stability in the
following form
\begin{equation}
\lambda_{n}-\lambda_{p}=2\frac{\partial (E/A)}{\partial X}
\biggr|_{A}=0\ .
\label{betastab}
\end{equation}
The beta-stability line $X^{\ast}(A)$ is directly derived from
Eqs.~(\ref{E/A}), (\ref{eC}) and (\ref{betastab}) as 
\begin{equation}
\quad X^{\ast}(A)=\frac{e_{C}(A)}{b(A)+e_{C}(A)}
\label{cond}
\end{equation}
We point out that for finite nuclei, the beta-stability condition
$\lambda_{n}-\lambda_{p}=0$ is not necessary fulfilled explicitly because of the
subshell structure in the discrete spectrum of the single particle levels
near the Fermi energy for both the neutrons and the protons. Note also that,
strictly speaking, the $\beta$-decay is forbidden if $\left|\lambda_{n}-
\lambda_{p}\right| < m_{e}c^{2}$, where $m_{e}$ is the
electron (positron) mass, i.e., in general, the condition $\lambda_{n}-\lambda _{p}=0$
for $\beta$-stability is too strong and we can expect more smooth behavior of
$X^{\ast}(A)$ than the one given by Eq.~(\ref{cond}).

Along the $\beta$-stability line, the binding energy per particle is given by 
\begin{equation}
E^{\ast}/A=e_{0}(A)+b(A) X^{\ast\,2}+E_{C}(X^{\ast})/A\ ,
\label{stab}
\end{equation}
where the upper index ``$\ast$'' indicates
that the corresponding quantity is determined by the variational conditions
(\ref{cond}) taken for fixed $A$ and $X=X^{\ast}$ on the beta-stability
line. For any given value of mass number $A$, the binding energy per nucleon 
$E/A$ can be extended beyond the beta-stability line as 
\begin{equation}
E/A=E^{\ast}/A+b(A)(X-X^{\ast})^{2}+\Delta E_{C}(X)/A\ ,
\label{eq1}
\end{equation}
where $\Delta E_{C}(X)=E_{C}(X)-E_{C}(X^{\ast})$. The symmetry energy
coefficient $b(A)$ contains the $A$-independent bulk term,
$b_{V}$, and the $A$-dependent surface contribution,
$b_{S}A^{-1/3}$, 
\begin{equation}
b(A)=b_{V}+b_{S}A^{-1/3}.
\label{bi}
\end{equation}
In general, the surface symmetry energy includes also the high order
curvature correction $\propto A^{-2/3}$ \cite{kosa08}.

Using Eq.~(\ref{eq1}), one can establish an important relation for the
chemical potentials $\lambda_{q}$ beyond the beta-stability line. Namely,
for the fixed particle number $A$, we obtain from Eqs.~(\ref{E/A}),
(\ref{stab}) and (\ref{cond}) the following relation 
\begin{equation}
\Delta\lambda (A,X)/4=(\lambda_{n}-\lambda_{p})/4
=\frac{1}{2}\frac{\partial (E/A)}{\partial X}\biggr|_{A}
=\left[ b(A)+e_{C}(A)\right] X-e_{C}(A)\ .
\label{lambdaX2}
\end{equation}
On the other hand, the shift $\Delta \lambda (A,X)$ of the neutron-proton
potentials can be evaluated numerically within the accuracy of $\sim 1/A^{2}$
using for the quantity of $\partial (E/A)/\partial X$ in Eq.~(\ref{lambdaX2})
the experimental values of the binding energy per nucleon
$\mathcal{B}(N,Z)=-E(N,Z)/A$. Namely, 
\begin{equation}
\frac{\partial (E/A)}{\partial X}\biggr|_{A}=\frac{A}{4}\left[\,
\mathcal{B}(N\!-\!1,Z\!+\!1)-\mathcal{B}(N\!+\!1,Z\!-\!1)\,\right]\ . 
\label{diff}
\end{equation}
Since the difference (\ref{diff}) is taken for $\Delta Z=-\Delta N=2$, the
pairing effects do not affect the resulting accuracy. It was shown in
Ref.~\cite{kosa10} that the linear dependence of $\Delta\lambda (A,X)$ given by
Eq.~(\ref{lambdaX2}) at fixed particle number $A=\mathrm{const}$ is
reproduced quite well experimentally. This fact allows one to extract the
values of $b(A)$, $e_{C}(A)$, and $X^{\ast}$ for a given
mass number $A$ with acceptable accuracy.

\begin{figure}
\includegraphics[width=0.48\textwidth]{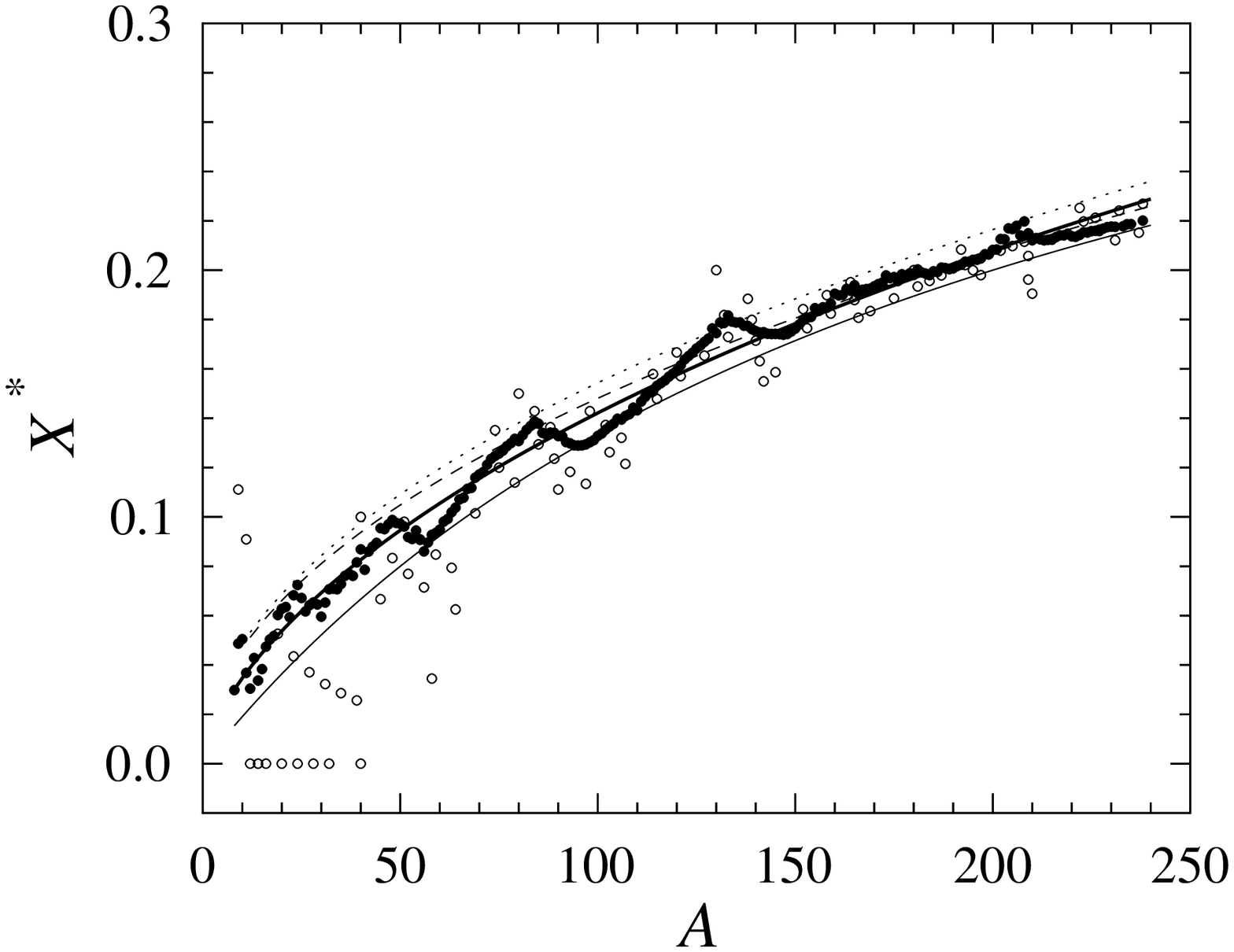}\hfill
\includegraphics[width=0.48\textwidth]{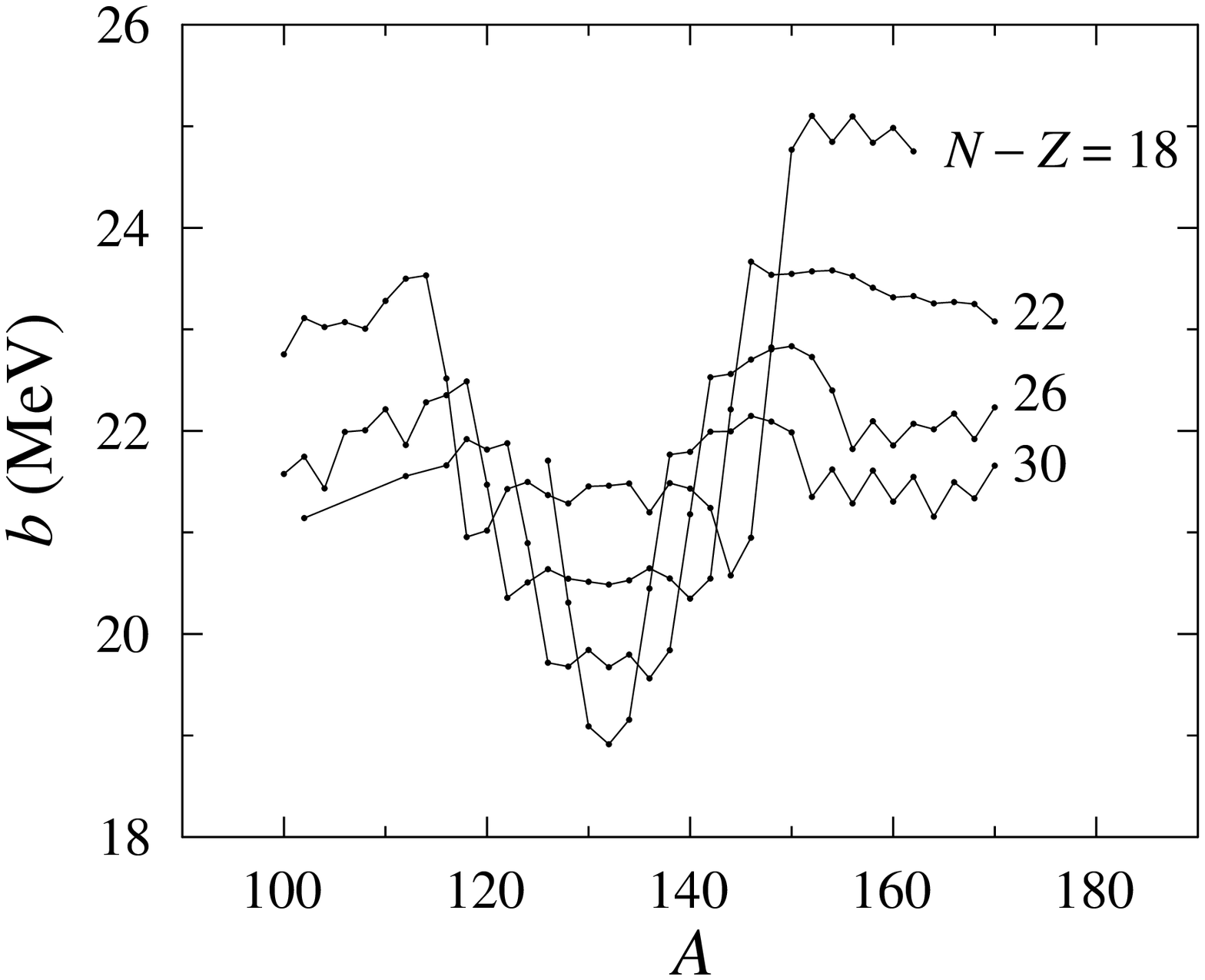}
\begin{minipage}[t]{0.48\textwidth}
\caption{Asymmetry parameter $X^{\ast}(A)$ versus the mass number $A$.
Solid dots represent the data obtained from the condition $\Delta\lambda (A,X)=0$.
Open circles correspond to the stable isotopes of maximum abundance for different
elements. Solid lines present $X^{\ast}(A)$ from Eq.~(\ref{xstar}) (thin)
%the functions $X^{\ast}(A)=0.4A/(A+200)$
%\cite{gren53}
and from Eq.~(\ref{xx}) with $b_{V}=27$~MeV,
$b_{S}=-23$~MeV (thick). The calculations using different Skyrme forces
are shown by the dashed (SLy230b) and dotted (SkM) lines \cite{kosa13}.\label{fig1}}
\end{minipage}\hfill
\begin{minipage}[t]{0.48\textwidth}
\caption{The symmetry coefficient $b$ vs mass number $A$ at fixed
neutron excess $A_{-}=N-Z$. The values of the neutron excess are specified by numbers
near the curves.\label{fig2}}
\end{minipage}
\end{figure}

Using Eqs. (\ref{lambdaX2}) and (\ref{diff}), we have evaluated the
"experimental" values of quantities $X^{\ast }(A)$ and $b(A)$
along the Periodic Table of the Elements. From the
beta-stability condition $\Delta\lambda (A,X)=0$ and Eqs.~(\ref{lambdaX2})
and (\ref{diff}) we can derive the asymmetry parameter $X^{\ast}(A)$. In
Fig.~\ref{fig1}, we have plotted the obtained "experimental" value of $X^{\ast}(A)$
(solid dots). The $\beta$-stability line $X^{\ast}(A)$ can be also evaluated
theoretically using an appropriate equation of state (EOS). In our numerical calculations
we have used the EOS from the extended Thomas-Fermi approximation (ETFA) with
Skyrme forces \cite{kosa13}.
The result of the typical microscopic calculation of $X^{\ast}(A)$ within
the extended Thomas-Fermi approximation
with Skyrme forces SLy230b is shown in Fig.~\ref{fig1}
as the dashed line. The numerical results presented in Fig.~\ref{fig1}
depends slightly only on the specific choice of Skyrme force
parametrization. For comparison the dotted line in Fig.~\ref{fig1}
shows the analogous result for Skyrme forces SkM.

The thin solid line in Fig.~\ref{fig1} was obtained by use the
phenomenological Green-Engler formula \cite{gren53} 
\begin{equation}
X^{\ast}(A)=\frac{0.4\,A}{A+200}\ .
\label{xstar}
\end{equation}
The "experimental" curve (solid dots) $X^{\ast}(A)$ in Fig.~\ref{fig1}
shows the non-monotonic (sawtooth) shape as a function of the mass number $A$.
This behavior is the consequence of subshell structure of the single
particle levels near the Fermi surface for both the neutrons and the
protons. Because of this subshell structure, the Fermi levels for protons
and neutrons can coincide (such a coincidence is the condition for the
$\beta $-stability) by chance only creating the non-monotonic behavior of
$X^{\ast}(A)$. Note that the non-monotonic subshell structure of the
$\beta$-stability line is transparently discovered for the curve $X^{\ast }(A)$
only, i.e., for $A$-dependency of $X^{\ast}$. The traditional
representation of $\beta$-stability line as $Z(N)$-dependency does not
allow one to observe this phenomenon. The reason is that the shell
oscillations appear against the small asymmetry parameter $X^{\ast}(A)$
which is close to zero. For the same reason the value of $X^{\ast}(A)$
requires more rigorous description than $Z(N)$.

We point out also that the traditionally used beta-stability line $Z(N)$ is
given for a discrete set of the asymmetry parameter $X$ and the
mass number $A$ which obey the condition $|\Delta\lambda(A,X)|< m_{e}c^{2}$.
Under this condition the
beta-stable nuclei represent rather eroded area than line as compared to
the more tight definition $\Delta\lambda (A,X)=0$. In Fig.~\ref{fig1}
we have plotted the discrete points of the beta-stability line $Z(N)$ as
the open circles. Each open circle in Fig.~\ref{fig1} corresponds
to the stable isotope of maximum abundance for a certain value of charge
number $Z$. As seen from Fig.~\ref{fig1}, there is a correlation
between the locations of solid dots and open circles. The location of the
$\beta$-stability line defined by the condition $\Delta\lambda (A,X)=0$
(solid dots) is obviously less scattered over the plot area, especially for
light nuclei. In practical sense, the $A$-dependent $\beta$-stability line
$X^{\ast}(A)$ is useful to extract the Coulomb energy
parameter $e_{C}(A)$ and the symmetry energy $b(A)$ from
the experimental data by use the chemical potential shifts
$\Delta\lambda (A,X)$, see e.g. Eq.~(\ref{lambdaX2}).

To show the origin of the subshell oscillations of $X^{\ast}(A)$ more
transparently, we will consider the sequence of the nucleon magic numbers 
\cite{bomo1}: 8, 20, 28, 50, 82 and 126. From this sequence one
should expect special behavior of the $X^{\ast}(A)$ nearby the following
values of mass number $A=N+Z$: 28\,(20+8), 48\,(28+20), 78\,(50+28),
132\,(82+50) and 208\,(126+82). The "experimental" beta-stability line
(solid dots in Fig.~\ref{fig1}) has the local maxima at mass
numbers 24\,(13+11), 48\,(26+22), 84\,(48+36), 133\,(79+54) and 208\,(126+82).
We can see that mass numbers of local maxima in Fig.~\ref{fig1}
does not exactly follow double magic numbers. Nevertheless, one can state
that, at least approximately, there exists the correlation between the
positions of maxima of sawtooth function $X^{\ast}(A)$ and double magic
mass numbers.

The Coulomb energy parameter $e_{C}(A)$ in Eq.~(\ref{lambdaX2}) can be
easily evaluated for a given proton density distribution independently on
the nuclear $NN$-interaction. In the simplest case, assuming a sharp proton
distribution and neglecting the contribution from the quantum exchange
term, one obtains $e_{C}(A)=0.15Ae^{2}/R_{C}\propto A^{2/3}$,
where $R_{C}$ is the charge (Coulomb) radius of nucleus. In general, both
the finite diffuse layer and the quantum exchange contributions must be
taken into account. The last fact leads to more complicate $A$-dependence of 
$e_{C}(A)$.
To extract such an actual $A$-dependency of the Coulomb parameter $e_{C}(A)$
which includes both above mentioned contributions, we will consider the
values of the chemical potential shift $\Delta\lambda (A,X)$ at the fixed
neutron excess, $A_{-}=N-Z=AX$ and the different particle numbers $A$. As seen
from Eq.~(\ref{lambdaX2}), for the zero's neutron excess $A_{-}=0$ the value of
$\Delta\lambda (A,X)$ is not affected by the symmetry energy $b$
and it is completely determined by $e_{C}(A)$. Due to this fact, for
nuclei with $A_{-}=0$ the Coulomb parameter $e_{C}(A)$ can be evaluated
precisely including all corrections caused by the finite diffuse layer, the
quantum exchange effects, etc. The Coulomb parameter $e_{C}(A)$ can be
represented by the smooth function 
\begin{equation}
e_{C}(A)=C_{1}A^{2/3}+C_{2}A^{1/3}
\label{ec}
\end{equation}
with $C_{1}=0.207$, $C_{2}=-0.174$ obtained using the fit to all available
"experimental" data with $A_{-}=0$.
%Note that the $A^{1/3}$-term in Eq.~(\ref{ec}) appears
%due to both the finite diffuse layer in proton density distribution and the
%quantum exchange correction to the full Coulomb energy.
The use of Eq. (\ref{lambdaX2}) for the shift $\Delta \lambda (X)$ at fixed
$A$ allows us to determine the Coulomb parameter $e_{C}(A)$ for the whole
region of mass number covered by experimental data. This was earlier done in
Ref.~\cite{kosa10} where the Coulomb parameter $e_{C}(A)$ was
roughly estimated as $e_{C}(A)\approx 0.17A^{2/3}$. However,
more precise evaluation is complicated because of the strong shell
oscillations at $e_{C}(A)$. In contrast, the data for $e_{C}(A)$
obtained from (\ref{lambdaX2}) at fixed $A_{-}=0$ do not
show much shell structure. This fact
is also supported by results of Ref. \cite{liwa11}.

Note that the actual value of the Coulomb parameter $e_{C}(A)$ can deviate
from its extrapolation given by Eq.~(\ref{ec}) for heavy nuclei with $X\neq 0$.
This deviation is caused by the fact that the proton distribution radius 
$R_{C}$ is slightly dependent on the neutron excess ("neutron skin") in
asymmetric nuclei. The origin of such dependency is the polarization effect.
Namely, the saturation bulk density decreases with $X$ for
neutron-rich nuclei where more neutrons are pushed off to the "neutron skin"
involving also the protons and increasing thereby the radius of proton
distribution. Such kind of polarization effect of the neutron excess on the
proton distribution can be estimated evaluating the $X$-dependency of the
bulk density in asymmetric nuclei \cite{kolu12,oyta98,oyii03}.
The estimation made in \cite{kosa13} shows that the influence of the neutron excess
on the Coulomb radius $R_{C}$ is negligible in asymmetric nuclei with $X\ll 1$ and the
extrapolation formula (\ref{ec}) for the Coulomb energy parameter $e_{C}(A)$
can be used with high accuracy for heavy nuclei with $X\neq 0$.

Taking into account Eqs.~(\ref{cond}), (\ref{bi}) and (\ref{ec}), we suggest
the following new form for the $\beta$-stability line 
\begin{equation}
X^{\ast}(A)=\frac{C_{1}A^{2/3}+C_{2}A^{1/3}}{C_{1}A^{2/3}+C_{2}A^{1/3}
+b_{V}+b_{S}A^{-1/3}}\ .
\label{xx}
\end{equation}
Fitting $X^{\ast}(A)$ in Fig.~\ref{fig1} by formula (\ref{xx}), we can
derive the smooth "experimental" parameters of the symmetry energy
$b_{V}$ and $b_{S}$. The corresponding smooth
behavior of $X^{\ast }(A)$ is shown in Fig.~\ref{fig1} by solid thick line.
This line was obtained as a best fit with the values of
$b_{\mathrm{sym,vol}}=27$~MeV and $b_{\mathrm{sym,surf}}=-23$~MeV which provide the
surface-to-volume ratio
$r_{S/V}=|b_{S}|/b_{V}\approx 0.85$.
Note that the analysis made in Ref.~\cite{bomo2} for the
saddle point shapes of fissile nuclei gives the value for the surface
symmetry coefficient of about $b_{S}\approx -25$~MeV.

%
% Figures 1,2
%

Approximating the contribution of the Coulomb energy to $\Delta\lambda (A,X)$
by Eq.~(\ref{ec}), one can extract $b(A)$ at fixed neutron
excess $A_{-}\neq 0$ from the experimental values of $\Delta\lambda$ by means
of Eq.~(\ref{lambdaX2}). We have performed such kind of numerical
calculations of the "experimental" symmetry energy coefficient $b(A)$
as a function of mass number $A$ beyond the $\beta$-stability
line for the values of the fixed neutron excess $A_{-}=18$, 22, 26
and 30. The corresponding results are shown in Fig.~\ref{fig2}.
As seen from Fig.~\ref{fig2}, qualitatively, $b(A)$
has canyon-like behavior for a given value of $A_{-}$. Such kind of canyon-like
behavior of the symmetry energy correlates with the nuclear subshell
structure. The width and the position of the bottom for the "canyon" depend
on the neutron excess $A_{-}$. The left wall of the canyon corresponds to the
proton closed shell and the right wall corresponds to the neutron shell
closure. Such kind of features can be understood from Eq.~(\ref{lambdaX2})
and the fact that the value of the nucleon chemical potential $\lambda_{q}$
goes up sharply when one moves from the closed shell to the one which is far
from closure. In Fig.~\ref{fig2} the walls are located
symmetrically with respect to $A=132$ which corresponds to both neutron and
proton closed shell ($N=82$, $Z=50$). From $A_{-}=18$ to $A_{-}=30$ the shape of
$b(A)$ changes to thinner and deeper canyon with the local
minimum in the symmetry coefficient being located at mass number which
corresponds to double (proton-neutron) magic number. One can conclude that
the thin canyon-like structure of the symmetry energy coefficient
$b(A)$ is caused by the shell effects in the single-particle
level distribution near the nucleon Fermi energy.

\section{ISOSPIN EFFECTS WITHIN GIBBS -- TOLMAN APPROACH}
We consider first the spherical nucleus at zero temperature, having the
mass number $A=N+Z$, the neutron excess $A_{-}=N-Z$ and the asymmetry
parameter $X=A_{-}/A$. The total binding energy of nucleus is $E$. An actual
nucleus has the finite diffuse layer of particle density distribution.
Thereby, the nuclear size is badly specified. In order to formulate proper
definition for the nuclear radius, we will use the concept of dividing
surface of radius $R$, originally introduced by Gibbs \cite{gibbs}.
Following Refs.~\cite{rowi82,gibbs}, we introduce the formal dividing
surface of radius $R$, the corresponding volume $V=4\pi R^{3}/3$
and the surface area $S=4\pi R^{2}$. Note that the dividing
surface is arbitrary but it should be located within the nuclear diffuse
layer.

The energy of a nucleus $E$, as well as the mass number $A$ and the neutron
excess $A_{-}$, are spitted into the volume and surface parts,
\begin{equation}
E=E_{V}+E_{S}\ +E_{C},\ \ \ \ 
%\end{equation}
%\begin{equation}
A=A_{V}+A_{S}\ ,\ \ \ 
A_{-}=A_{-,V}+A_{-,S}.  \label{parts}
\end{equation}
Here the Coulomb energy $E_{C}$ is fixed and does not depend on the dividing
radius $R$. The bulk energy $E_{V}$ and the surface energies
$E_{S}$ can be written as \cite{lali58,rowi82}
\begin{equation}
E_{V}=\left( -P+\lambda\varrho_{V}+
\lambda_{-}\varrho_{-,V}\right)V\ \ \ 
%\end{equation}
\mbox{and\ \ \ }
%\begin{equation}
E_{S}=\left(\sigma +\lambda\varrho_{S}+
\lambda_{-}\varrho_{-,S}\right)S.  \label{surfvol}
\end{equation}
Here $P$ is the bulk pressure
\begin{equation}
P=-\left.\frac{\partial E_{V}}{\partial V}
\right\vert_{A_{V}},  \label{pv}
\end{equation}
$\sigma$ is the surface tension and
$\varrho_{V}=A_{V}/V$ and
$\varrho_{-,V}=A_{-,V}/V$
are, respectively, the total (isoscalar) and the neutron excess (isovector)
volume densities, $\varrho_{S}=A_{S}/S$ and
$\varrho_{-,S}=A_{-,S}/S$ are the
corresponding surface densities. We have used the isoscalar
$\lambda=(\lambda_{n}+\lambda _{p})/2$ and isovector
$\lambda _{-}=(\lambda_{n}-\lambda _{p})/2$ chemical potentials,
where $\lambda _{n}$ and $\lambda_{p}$ are the chemical potentials
of neutron and proton, respectively. The Coulomb energy $E_{C}$ must be excluded
from the chemical potentials $\lambda$ and $\lambda_{-}$ because of
Eqs.~(\ref{parts}) and (\ref{surfvol}). Namely, 
\begin{equation}
\lambda_{n}=\left.\frac{\partial E}{\partial N}\right\vert_{Z},\quad
\lambda_{p}=\left.\frac{\partial E}{\partial Z}\right\vert_{N}-
\lambda_{C}\ ,\mbox{\ \ \ \ where\ \ }
\lambda_{C}=\left.\frac{\partial E_{C}}{\partial Z}\right\vert_{N}. 
\label{lambda}
\end{equation}
%where
%\[
%\lambda_{C}=\left.\frac{\partial E_{C}}{\partial Z}\right\vert_{N}. 
%\]
Note that the definition of $\lambda_{p}$ in Eq.~(\ref{lambda}) differs
from the previous one given by (\ref{dlamb}).
Notation $E_{V}$ stands for the nuclear matter energy of the
uniform densities $\varrho_{V}$, $\varrho_{-,V}$
within the volume $V$. The state of the nuclear matter
inside the specified volume $V$ is chosen to have the chemical
potentials $\mu$ and $\mu_{-}$ equal to that of the actual droplet. In
more detail, from the equation of state for the nuclear matter one has
chemical potentials $\mu (\rho ,\rho_{-})$ and $\mu_{-}(\rho ,\rho_{-})$
as functions of the isoscalar, $\rho$, and isovector, $\rho_{-}$,
densities. Then, the following conditions should be fulfilled:
\begin{equation}
\mu (\rho =\varrho_{V},\ \rho_{-}=
\varrho_{-,V})=\lambda\ ,\ \ \ \ 
\mu_{-}(\rho =\varrho_{V},\ \rho_{-}=
\varrho_{-,V})=\lambda_{-}   \label{matter}
\end{equation}
to derive the specific values of densities $\varrho_{V}$ and 
$\varrho_{-,V}$.

The surface part of the energy $E_{S}$ as well as the surface
particle number $A_{S}$ and the surface neutron excess
$A_{-,S}$ are considered as the excess quantities responsible for
``edge'' effects with respect to the corresponding volume quantities.
Using Eqs.~(\ref{parts}), (\ref{surfvol}) one obtains 
\begin{equation}
\sigma =\frac{E-\lambda A-\lambda_{-}A_{-}}{S}+
\frac{P V}{S}-
\frac{E_{C}}{S}=
\frac{\Omega -\Omega_{V}}{S}\ .\label{sigma0}
\end{equation}
Here the grand potential $\Omega =E-\lambda A-\lambda_{-}A_{-}-E_{C}$ and
its volume part $\Omega_{V}=-P V=
E_{V}-\lambda A_{V}-\lambda_{-}A_{-,V}$ were
introduced. From Eq.~(\ref{sigma0}) one can see how the value of the surface
tension depends on the choice of the dividing radius $R$, 
\begin{equation}
\sigma\left[ R\right] =\frac{\Omega}{4\pi R^{2}}+
\frac{1}{3}P R\ .  \label{sigmaR}
\end{equation}

Taking the derivative from Eq.~(\ref{sigmaR}) with respect to the formal
dividing radius $R$ and using the fact that observables $E$, $\lambda $,
$\lambda_{-}$ and $P$ should not depend on the choice of the dividing
radius, one can rewrite Eq.~(\ref{sigmaR}) as 
\begin{equation}
P=2\,\frac{\sigma\left[ R\right]}{R}+
\frac{\partial }{\partial R}\,\sigma\left[ R\right]\ ,  \label{genlap}
\end{equation}
which is the generalized Laplace equation. The formal values of surface
densities $\varrho_{S}$ and $\varrho_{-,S}$ can be
found from (\ref{parts}) as
\begin{equation}
\varrho_{S}[R]=
\frac{A}{4\pi R^{2}}-\frac{1}{3}\varrho_{V}R\ ,\ \ \ \ 
\varrho_{-,S}[R]=
\frac{A_{-}}{4\pi R^{2}}-\frac{1}{3}\varrho_{-,V}R\ .\label{surfden}
\end{equation}
In Eqs.~(\ref{sigmaR}) -- (\ref{surfden}) square brackets denote a formal
dependence on the dividing radius $R$ which is still arbitrary and may not
correspond to the actual physical size of the nucleus. To derive the
physical size quantity an additional condition should be imposed on the
location of dividing surface. In general, the surface energy $E_{S}$
for the arbitrary dividing surface includes the contributions from the
surface tension $\sigma$ and from the binding energy of particles within
the surface layer. The latter contribution can be excluded for the special
choice of dividing (equimolar) radius $R=R_{e}$ which satisfy the condition 
\begin{equation}
\left(\varrho_{S}\lambda +
\varrho _{-,S}\lambda_{-}\right)_{R=R_{e}}=0\ .  \label{emolar}
\end{equation}
Here we use the notation $R_{e}$ by the analogy with the equimolar dividing
surface for the case of the one-component liquid \cite{kosa12,rowi82}. For
the dividing radius defined by Eq.~(\ref{emolar}) the surface energy reads 
\begin{equation}
E_{S}=\sigma_{e}S_{e}\ ,  \label{efree}
\end{equation}
where $\sigma _{e}\equiv\sigma (R_{e})$ and $S_{e}=4\pi R_{e}^{2}$.
Using Eqs.~(\ref{surfden}), (\ref{emolar}), the corresponding volume
$V_{e}=4\pi R_{e}^{3}/3$ is written as 
\begin{equation}
V_{e}=\frac{\lambda A+\lambda_{-}A_{-}}
{\lambda \varrho_{V}+\lambda _{-}\varrho _{-,V}}\ .
\label{evol}
\end{equation}
As seen from Eqs.~(\ref{matter}), (\ref{evol}), the droplet radius $R_{e}$
is determined by the equation of state for the nuclear matter through the
values of the droplet chemical potentials $\lambda$ and $\lambda_{-}$.

\begin{figure}
\includegraphics[width=0.48\textwidth]{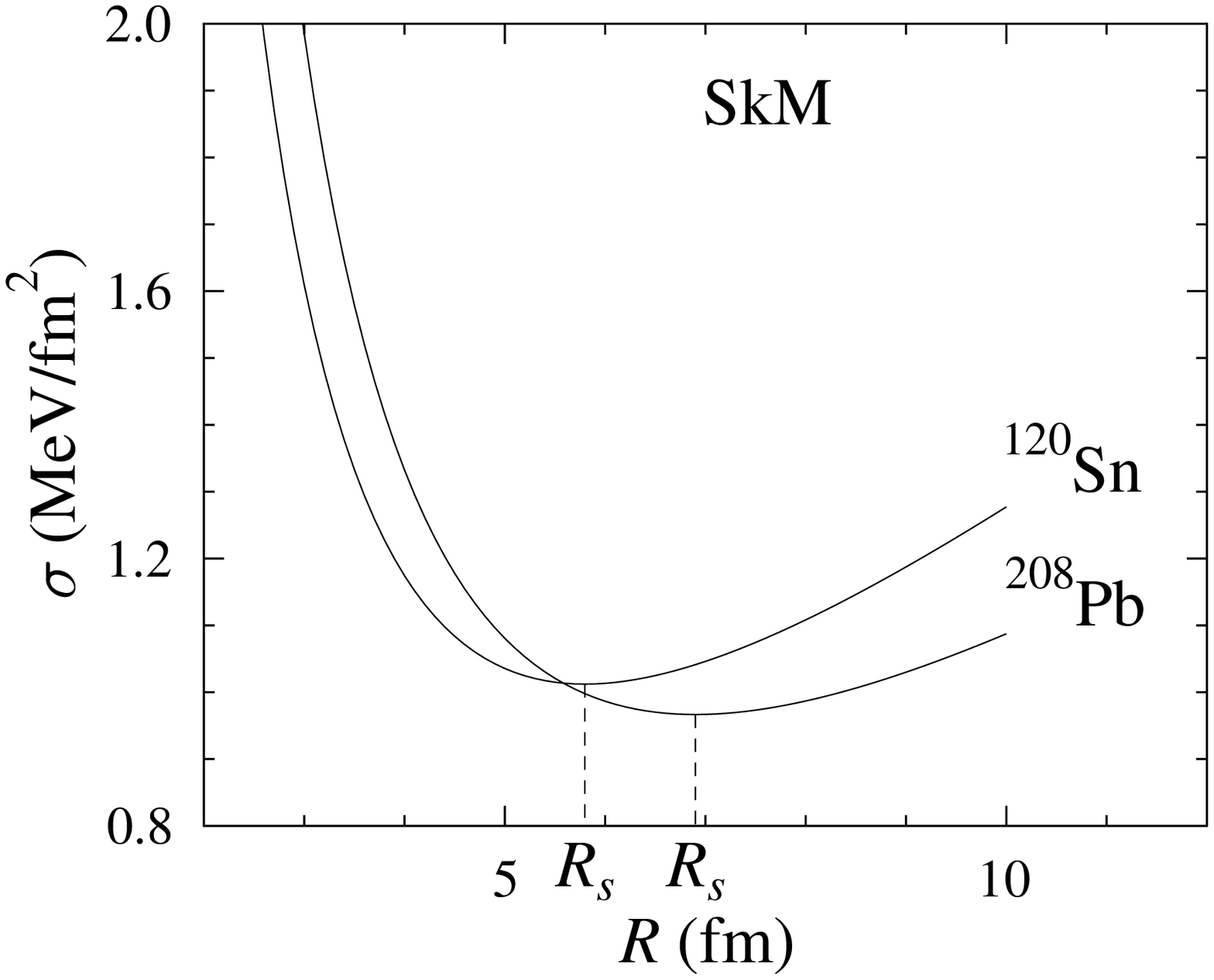}\hfill
\includegraphics[width=0.48\textwidth]{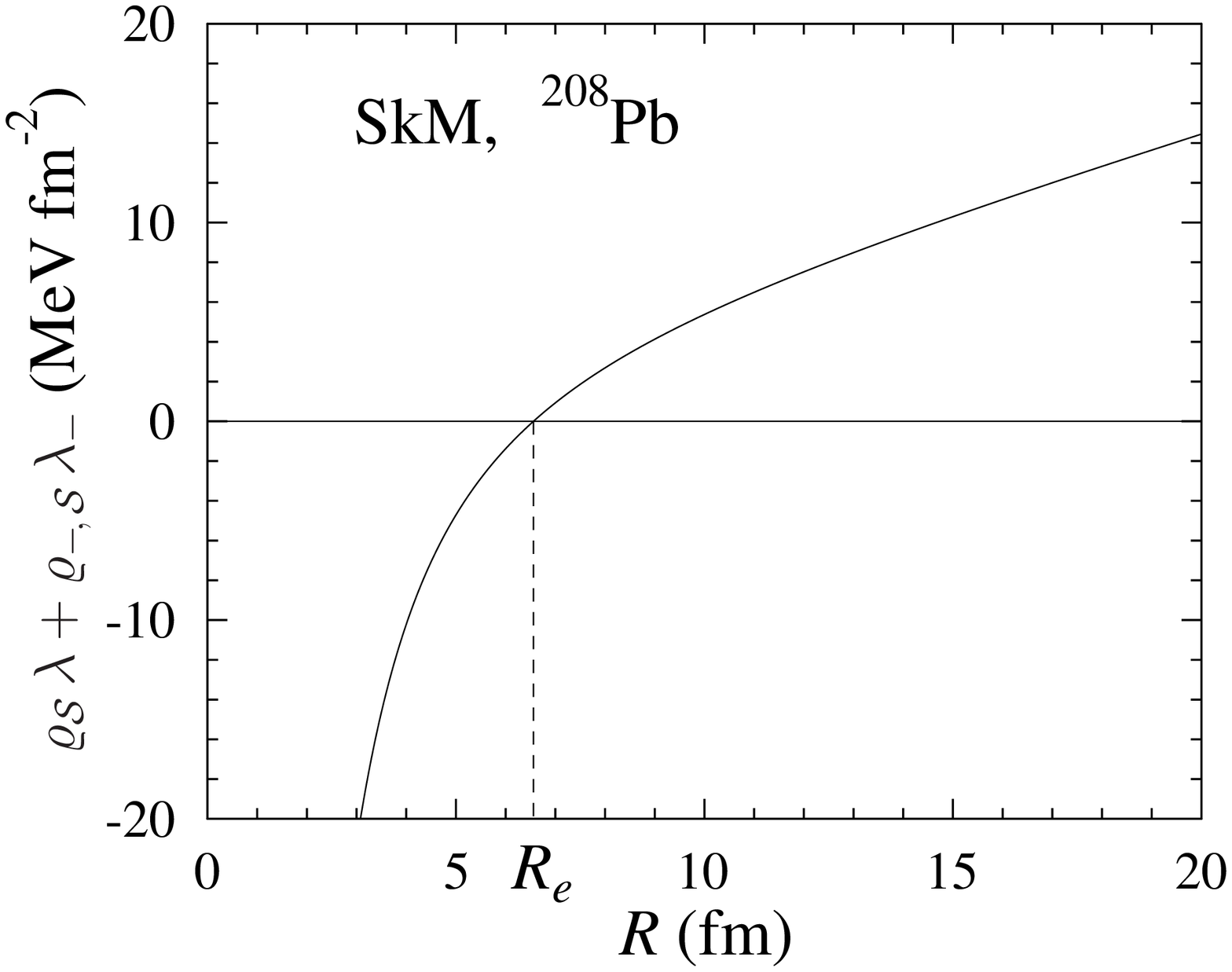}
\begin{minipage}[t]{0.48\textwidth}
\caption{
Surface tension $\sigma$ as a function of the dividing
radius $R$ for nuclei $^{120}$Sn and $^{208}$Pb. The calculation was
performed using SkM force (see \cite{kosa13a} for details).
The Laplace radius $R_{s}$ denotes the dividing radius where
$\sigma$ approaches the minimum value, i.e., the Laplace condition
of Eq. (\ref{p3}) is satisfied.
\label{fig3}}
\end{minipage}\hfill
\begin{minipage}[t]{0.48\textwidth}
\caption{
Specific surface particle density $\varrho_{S}\lambda +\varrho_{-,S}\lambda_{-}$
versus dividing radius $R$ for $^{208}$Pb.
The calculation was performed using the SkM force. $R_{e}$ denotes the
equimolar radius where $\varrho_{S}\lambda +\varrho_{-,S}\lambda_{-}$ becomes zero.
\label{fig4}}
\end{minipage}
\end{figure}

The surface tension $\sigma\left[ R\right]$ depends on the location of the
dividing surface. Function $\sigma\left[ R\right] $ has a minimum at
certain radius $R=R_{s}$ (radius of the surface of tension \cite{rowi82})
which usually does not coincide with the equimolar radius $R_{e}$. The
radius $R_{s}$ (Laplace radius) denotes the location within the interface.
Note that for $R=R_{s}$ the capillary pressure of Eq.~(\ref{genlap})
satisfies the classical Laplace relation 
\begin{equation}
P=2\left.\frac{\sigma\left[ R\right]}{R}
\right\vert_{R=R_{s}}\ .  \label{p3}
\end{equation}
The dependence of the surface tension $\sigma\left[ R\right]$ of
Eq.~(\ref{sigmaR}) on the location of the dividing surface for the
nuclei $^{120}$Sn and $^{208}$Pb is shown in Fig.~\ref{fig3}.

Following Gibbs and Tolman \cite{gibbs,tolm49}, we will assume that the
physical (measurable) value of the surface tension is that taken
at the equimolar dividing surface. We assume, see also Ref.~\cite{rowi82},
that the surface tension $\sigma\equiv\sigma (R_{e})$ approaches the
planar limit $\sigma_{\infty }$ as 
\begin{equation}
\sigma (R_{e})=\sigma_{\infty}\left( 1-\frac{2\xi}{R_{e}}+
\mathcal{O}(R_{e}^{-2})\right)\ ,  \label{sigmaeq}
\end{equation}
where $\xi$ is the Tolman's length \cite{tolm49}.
Taking Eq.~(\ref{genlap}) for $R=R_{s}$ and comparing with analogous one for 
$R=R_{e}$, one can establish the following important relation
(see Eq.~(\ref{R_e_s}) in the next Section)
\begin{equation}
\xi =\lim_{A\rightarrow\infty}({R_{e}-R_{s})}\ +\mathcal{O}(X^{2}).
\label{ksi1}
\end{equation}
This result leads to the conclusion that to obtain the non-zero value of
Tolman length $\xi$, and, consequently, the curvature correction
$\Delta\sigma_{\mathrm{curv}}\neq 0$ for a curved surface,
the nucleus must have a finite diffuse surface layer.

We perform the numerical calculations using Skyrme type of the
effective nucleon-nucleon interaction. The energy and the chemical
potentials for actual droplets can be calculated using a direct variational
method within the extended Thomas-Fermi approximation \cite{kosa08,kosa13a}.
Using obtained chemical potentials we evaluate the equilibrium bulk densities
$\varrho_{V}$ and $\varrho_{-,V}$ from Eq.~(\ref{matter}).
For arbitrary dividing radius $R$ and fixed asymmetry parameter $X$ we
evaluate then the volume, $A_{V}=4\pi\varrho_{V}R^{3}/3$
and $A_{-,V}=4\pi\varrho_{-,V}R^{3}/3$, the
surface, $A_{S}=4\pi\varrho_{S}R^{2}$ and
$A_{-,S}=4\pi\varrho_{-,S}R^{2}$, particle numbers and the
volume part of equilibrium energy $E_{V}$. All evaluated values of 
$E_{V}[R]$, the bulk densities $\varrho_{V}$ and
$\varrho_{-,V}$ and the surface particle densities
$\varrho_{S}[R]$ and $\varrho_{-,S}[R]$ depend on the radius $R$
of dividing surface and asymmetry parameter $X$. The actual physical radius
$R_{e}$ of the droplet can be derived by the condition (\ref{emolar}), i.e.,
by the requirements that the contribution to $E_{S}$ from the bulk
binding energy (term $\sim (\varrho _{S}\lambda +\varrho_{-,S}\lambda _{-})$
in Eq. (\ref{surfvol})) should be excluded from
the surface energy $E_{S}$.
In Fig.~\ref{fig4} we represent the calculation of the specific
surface particle density $\varrho_{S}\lambda +\varrho_{-,S}\lambda_{-}$
as a function of the radius $R$ of dividing
surface. Equimolar dividing radius $R_{e}$ in Fig.~\ref{fig4} defines the
physical size of the sharp surface droplet and the surface at which the
surface tension is applied, i.e., the equimolar surface where Eq. (\ref{efree})
is fulfilled.
%
% Figures 3,4
%

\begin{figure}
\begin{minipage}{0.48\textwidth}
\centerline{\includegraphics[width=\textwidth]{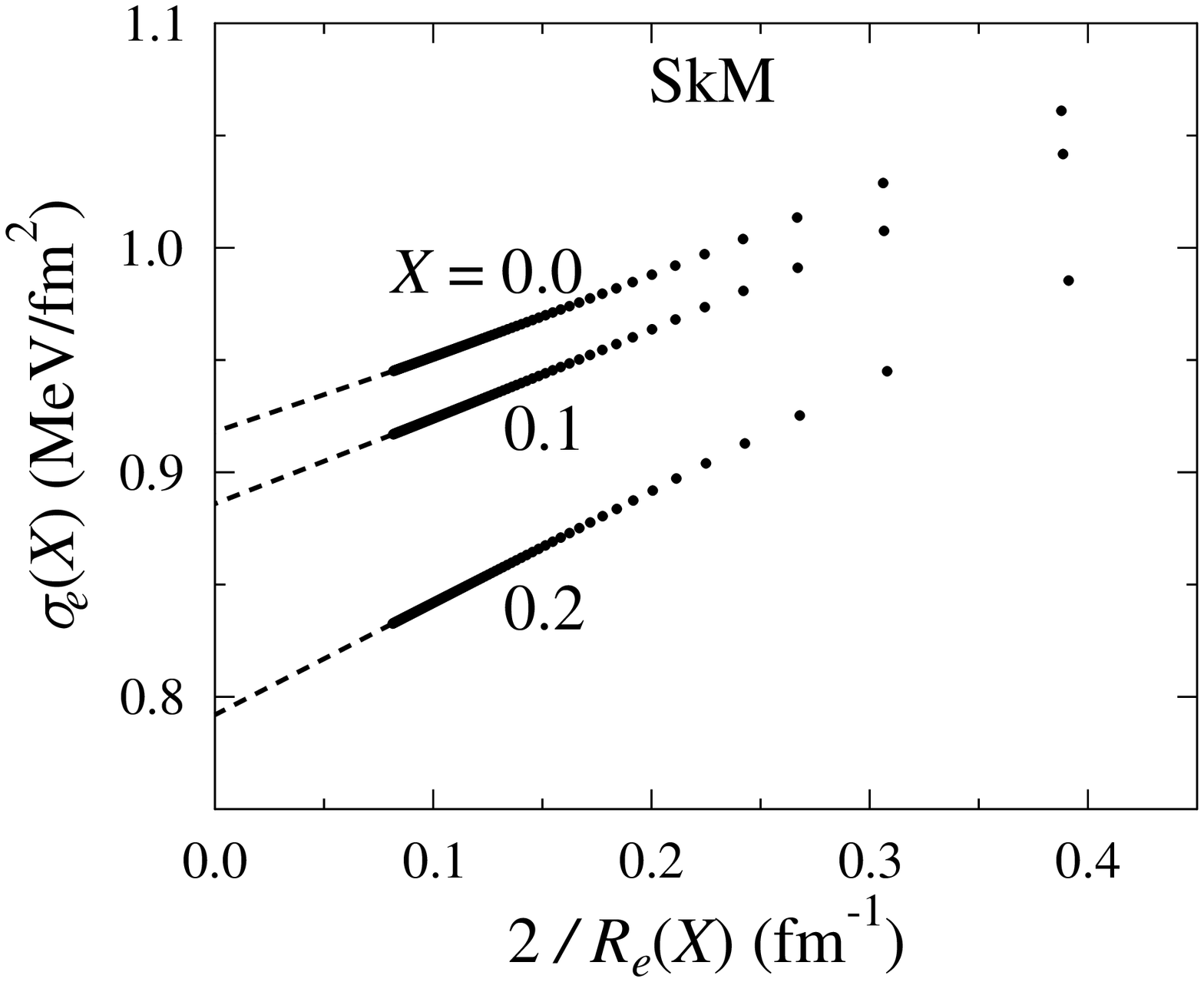}}
\end{minipage}\hfill
\begin{minipage}{0.48\textwidth}
\caption{
The dependence of the surface tension coefficient $\sigma\left(R_{e},X\right)$
on the equimolar radius $R_{e}$ for different values of the asymmetry parameter
$X$. The calculation was performed for Skyrme force SkM \cite{kosa13a}.
\label{fig5}}
\end{minipage}
\end{figure}

Note that the value of equimolar radius $R_{e}$, which is derived by Eq.~(\ref{evol}),
is not considerably affected by the Coulomb interaction. We
have also evaluated the values of $R_{e}$ neglecting the Coulomb term,
i.e., assuming $E_{C}=\lambda _{C}=0$. The difference as
compared with data obtained with Coulomb term included
does not exceed 0.5\% for $A$ of about 200. Omitting the
Coulomb energy contribution to the total energy $E$
and evaluating the bulk energy $E_{V}$, one can
obtain the surface part of energy $E_{S}=E-E_{V}$ and
the surface tension coefficient $\sigma\left( R_{e}\right)$ (\ref{sigma0})
at the equimolar dividing surface for nuclei with different mass number
$A\propto R_{e}^{3}$ and asymmetry parameter $X$. The dependence of the surface
tension coefficient $\sigma\left( R_{e}\right)$ on the doubled inverse
equimolar radius $2/R_{e}$ (see Eq. (\ref{sigmaeq})) is shown in Fig.~\ref{fig5}.
%
% Figure 5
%

\begin{figure}
\includegraphics[width=0.48\textwidth]{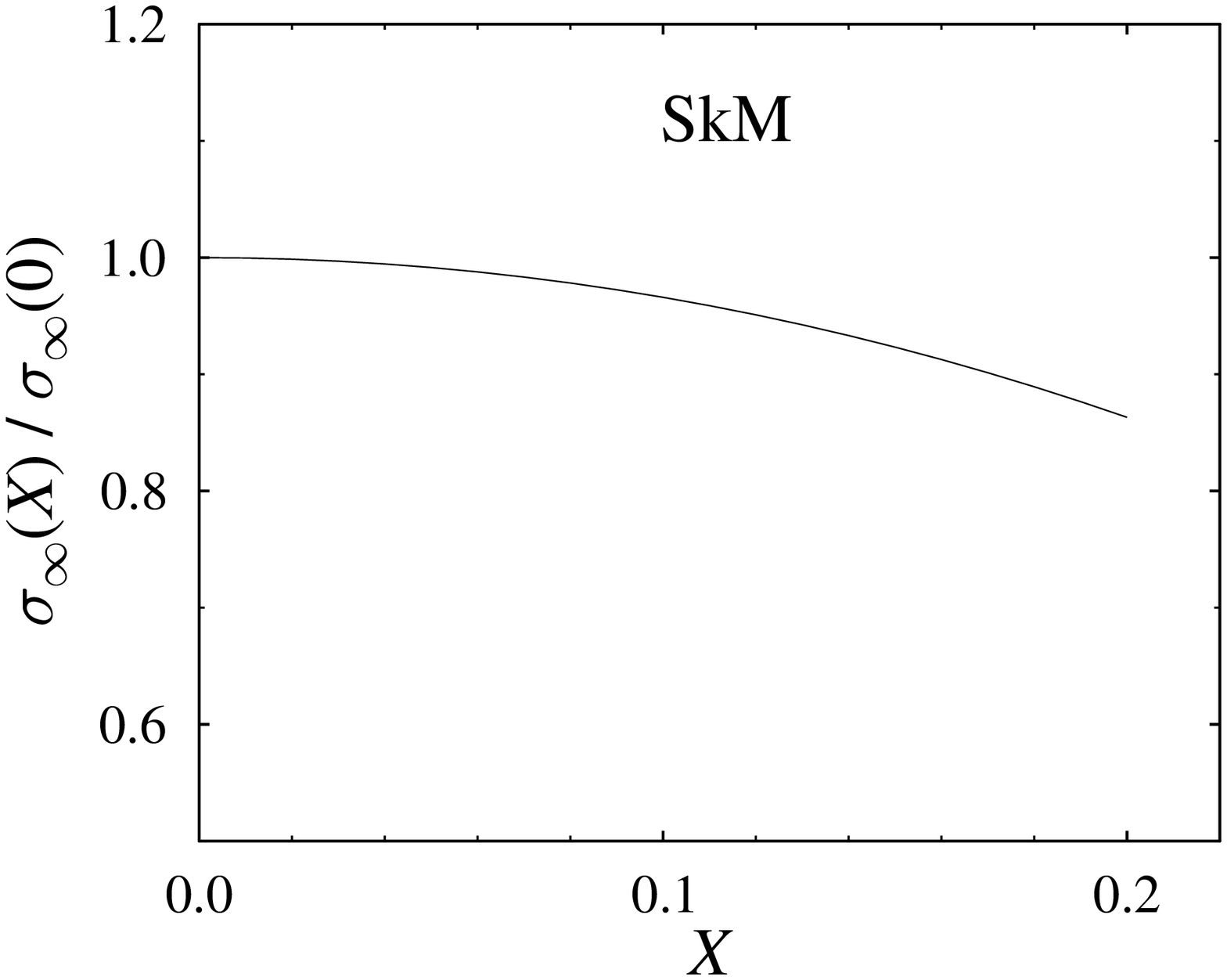}\hfill
\includegraphics[width=0.48\textwidth]{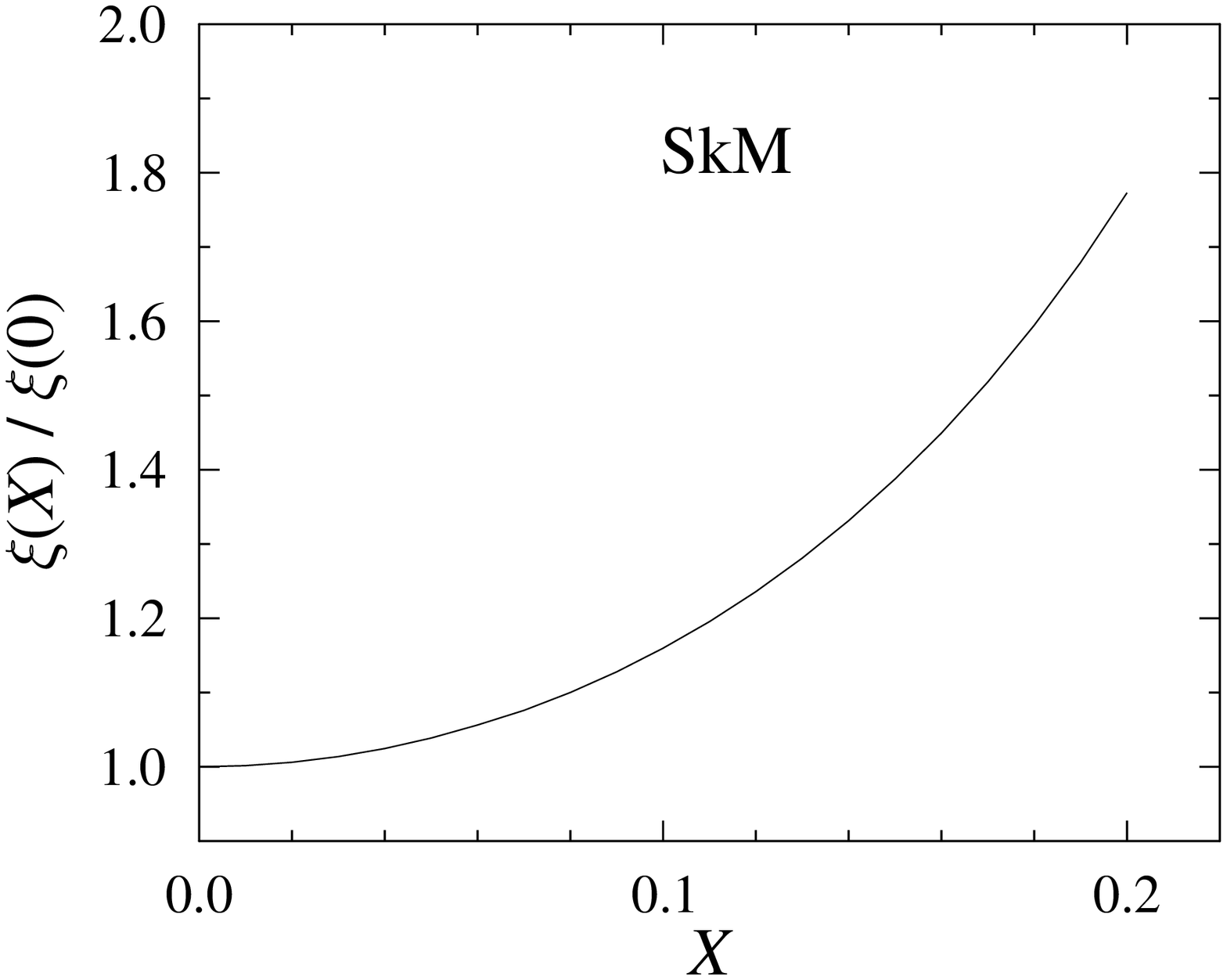}
\begin{minipage}[t]{0.48\textwidth}
\caption{
Dependence of the planar surface tension $\sigma_{\infty}(X)$
on the asymmetry parameter $X$. The calculation was
performed for Skyrme force SkM.
\label{fig6}}
\end{minipage}\hfill
\begin{minipage}[t]{0.48\textwidth}
\caption{
Dependence of the Tolman length $\xi$ on the asymmetry parameter $X$. The
calculation was performed for Skyrme force SkM.
\label{fig7}}
\end{minipage}
\end{figure}

The surface tension $\sigma\left(R_{e},X\right) $ approaches the planar
limit $\sigma_{\infty }(X)$ in the limit of zero curvature $2/R_{e}\rightarrow 0$.
As seen from Fig.~\ref{fig5}, the planar limit $\sigma_{\infty}(X)$ depends on the
asymmetry parameter. This dependence
reflects the fact that the symmetry energy $b$ in mass formula contains both
the volume $b_{V}$ and surface $b_{S}$ contributions, see Refs. \cite{dani03,kosa10}.
In Fig.~\ref{fig6} we show the $X$-dependence of the surface tension
$\sigma_{\infty}(X)$. This dependence can be approximated by
\begin{equation}
\sigma_{\infty}(X)=\sigma_{0}+\sigma_{-}X^{2}\ .  \label{sigma1}
\end{equation}
The dependence of parameters $\sigma_{0}$ and $\sigma_{-}$ on the Skyrme
force parametrization is shown in Table~\ref{tab1}. The isovector term $\sigma_{-}$
in the surface tension (\ref{sigma1}) is related to the surface contribution $b_{S}$
to the symmetry energy (see the next Section, Eq.~(\ref{as1x})).
We evaluate the surface-to-volume ratio
$r_{S/V}=|b_{S}/b_{V}| =1.17\div 1.47$ for Skyrme
force parametrizations from Table~\ref{tab1}. Note that in the previous
theoretical calculations, the value of surface-to-volume ratio $r_{S/V}$
varies strongly within the interval $1.6\leq r_{S/V}\leq 2.8$, see
Refs.~\cite{dani03,kosa10,sawy06}.

The slope of curves $\sigma\left( R_{e}\right)$ in Fig.~\ref{fig5} gives
the Tolman length $\xi$, see Eq. (\ref{sigmaeq}). The value of the Tolman
length $\xi$ depends significantly on the asymmetry parameter $X$. In 
Fig.~\ref{fig7} we show such kind of dependence obtained from results of 
Fig.~\ref{fig5}.
As seen from Fig.~\ref{fig7}, one can expect the enhancement of the
curvature effects in neutron rich nuclei. The $X$-dependence of Tolman
length $\xi$ can be approximated as 
\begin{equation}
\xi (X)=\xi_{0}+\xi_{-}X^{2}\ .  \label{xsi1}
\end{equation}%
Both parameters $\xi_{0}$ and $\xi_{-}$ as well as the surface tension
parameter $\sigma_{-}$ are rather sensitive to the Skyrme force
parametrization, see Table~\ref{tab1}.
%
% Figures 6,7
%

\section{NUCLEAR MATTER EQUATION OF STATE AND ($A^{-1/3}$, $X$)-EXPANSION
FOR FINITE NUCLEI}
We will consider the relation of the nuclear macroscopic
characteristics (surface and symmetry energies, Tolman length,
incompressibility, etc.) to the bulk properties of nuclear matter. Assuming
a small deviations from the equilibrium, the equation of state (EOS) for an
asymmetric nuclear matter can be written in the form expansion around the
saturation point. One has for the energy per particle (at zero
temperature) 
\begin{equation}
\mathcal{E}(\epsilon ,x)=\frac{E_{\infty}}{A}=
\mu_{\infty}+\frac{K_{\infty}}{18}\epsilon^{2}+
b_{\infty}x^{2}+\ldots\ ,  \label{fmatter}
\end{equation}
where
\[
\epsilon =\frac{\rho-\rho_{\infty}}{\rho_{\infty}}\ ,
\  x=\frac{\rho_{-}}{\rho}\ ,
\ \rho =\rho_{n}+\rho_{p}\ ,
\ \rho_{-}=\rho_{n}-\rho_{p}\ ,
\]
$\rho_{\infty}$ is the matter saturation (equilibrium) density,
$\mu_{\infty }$ is the chemical potential, $K_{\infty }$ is the nuclear matter
incompressibility and $b_{\infty}$ is the symmetry energy coefficient (all
values are taken at the saturation point $\epsilon=0$ and $x=0$).
Coefficients of expansion (\ref{fmatter}) are determined through the
derivatives of the energy per particle $\mathcal{E}(\epsilon ,x)$ at the
saturation point:
\begin{equation}
\mu_{\infty}=\left.\frac{E_{\infty }}{A}
\right\vert_{\rm s.p.}
%{\rho =\rho_{\infty},\,x=0}
\equiv\mathcal{E}^{(0,0)}\ ,\ \ \ \ 
K_{\infty}=9\left.\rho^{2}\frac{\partial^{2}E_{\infty}/A}
{\partial\rho^{2}}
\right\vert_{\rm s.p.}
%{\rho=\rho_{\infty},\,x=0}
\equiv
9\,\mathcal{E}^{(2,0)}\ ,\ \ \ \ 
%\end{equation}
%\begin{equation}
b_{\infty}=\frac{1}{2}\left.\frac{\partial^{2}E_{\infty}/A}
{\partial x^{2}}
\right\vert_{\rm s.p.}
%{\rho=\rho_{\infty},\,x=0}
\equiv\frac{1}{2}
\mathcal{E}^{(0,2)}\ .  \label{mKJ}
\end{equation}
Here we use the short notations
${\rm s.p.}\equiv(\rho=\rho_{\infty},\,x=0)$ and
%\[
$\mathcal{E}^{(n,m)}\equiv{\displaystyle\left.\frac{\partial^{n+m}\mathcal{E}}
{\partial\epsilon^{n}\partial x^{m}}\right\vert_{\epsilon=0,\,x=0}}$.
%\]
Some coefficients $\mathcal{E}^{(n,m)}$ are vanishing. From the condition of
minimum of $\mathcal{E}(\epsilon ,x)$ at the saturation point one has
$\mathcal{E}^{(1,0)}=\mathcal{E}^{(0,1)}=0$. Odd derivatives with respect to
$x$, i.e., $\mathcal{E}^{(n,m)}$ for odd $m$, also vanish because of the
charge symmetry of nuclear forces.
Using $\mathcal{E}(\epsilon ,x)$, one can also evaluate chemical
potentials $\mu$, $\mu_{-}$ and pressure $P$ of the nuclear matter beyond
the saturation point. Namely,
\begin{equation}
\mu (\epsilon ,x) = \left.\frac{\partial E_{\infty }}{\partial A}
\right\vert_{A_{-},V}=
\frac{\partial}{\partial\epsilon}(1+\epsilon )\mathcal{E}-
x\frac{\partial\mathcal{E}}{\partial x}\ ,\ \ \ \
\mu_{-}(\epsilon ,x)=\left.\frac{\partial E_{\infty }}{\partial A_{-}}
\right\vert_{A,V}=\frac{\partial\mathcal{E}}{\partial x}\ ,\label{mu01p} 
\end{equation}
\begin{equation}
P(\epsilon ,x) =-\left.\frac{\partial E_{\infty}}{\partial V}
\right\vert_{A,A_{-}}=\rho_{\infty}(1+\epsilon )^{2}
\frac{\partial\mathcal{E}}{\partial\epsilon}\ .  \label{pex1}
\end{equation}

Similarly to Eq.~(\ref{fmatter}), in a finite uncharged system the energy
per particle $E/A$ (we use $A=N+Z$, $A_{-}=N-Z$, $X=A_{-}/A$) of the
finite droplet is usually presented as ($A^{-1/3}$, $X$)-expansion around
infinite matter using the leptodermous approximation 
%\begin{equation}
%E/A=a_{V}+X^{2}b_{V}+A^{-1/3}\left(a_{S}+X^{2}\,b_{S}\right)+
%A^{-2/3}\left(a_{c}+X^{2}\,b_{c}\right) \label{eldm1}
%\end{equation}
\begin{equation}
E/A=a_{V}+a_{S}A^{-1/3}+a_{c}A^{-2/3}+
X^{2}(b_{V}+b_{S}A^{-1/3}+b_{c}A^{-2/3})
%\label{eldm2}
\label{eldm1}
\end{equation}
where $a_{V}$, $a_{S}$ and $a_{c}$ are, respectively, the volume, surface
and curvature energy coefficients, $b_{V}$, $b_{S}$ and $b_{c}$ are,
respectively, the volume, surface and curvature symmetry coefficients. The
nuclear chemical potentials $\lambda$ and $\lambda_{-}$ are derived as
\begin{equation}
\lambda (X,A^{-1/3})=E/A-\frac{1}{3}A^{-1/3}\frac{\partial\,E/A}{\partial A^{-1/3}}-
X\frac{\partial \,E/A}{\partial X}\ ,\ \ \ \ 
\lambda_{-}(X,A^{-1/3})=\frac{\partial\,E/A}{\partial X}\ .  \label{lambda01}
\end{equation}
Following Gibbs-Tolman method, one can derive the actual nuclear matter
densities $\rho$ and $\rho_{-}$ from the conditions 
\begin{equation}
\mu (\epsilon ,x)=\lambda (X,A^{-1/3})\ ,\ \ \ \ 
\mu_{-}(\epsilon ,x)=\lambda_{-}(X,A^{-1/3})\ .  \label{eosmatt}
\end{equation}
Using Eq.~(\ref{eosmatt}), one can establish the relation of the macroscopic
energy coefficients in the mass formula expansion Eq.~(\ref{eldm1}) to
the nuclear matter parameters in EOS (\ref{fmatter}),
see Eqs.~(\ref{av1x}) -- (\ref{R_e_s}) below.
The results of numerical calculations
of relevant quantities are represented in Tables~\ref{tab1} and \ref{tab2}.

We start from the nuclear matter EOS given by Eq. (\ref{fmatter}) and
take into consideration the relations (\ref{mKJ}) and the
following higher order coefficients 
\begin{equation}
K_{3}=6K_{\infty}+27\left.\rho^{3}\frac{\partial^{3}E_{\infty}/A}
{\partial\rho^{3}}\right\vert_{\rm s.p.}
%{\rho=\rho_{\infty},\,x=0}
\equiv 27\left(\mathcal{E}^{(3,0)}+2\,\mathcal{E}^{(2,0)}\right)\ ,\ \ \ \ 
L_{\infty}=\frac{3}{2}\left.\rho\frac{\partial^{3}E_{\infty}/A}
{\partial\rho\partial x^{2}}\right\vert_{\rm s.p.}
%{\rho=\rho_{\infty},\,x=0}
\equiv\frac{3}{2}\,\mathcal{E}^{(1,2)}\ ,
\label{homom}
\end{equation}
\begin{equation}
K_{\mathrm{sym}}=\frac{9}{2}\left.\rho^{2}\frac{\partial^{4}E_{\infty}/A}
{\partial\rho^{2}\partial x^{2}}\right\vert_{\rm s.p.}
%{\rho=\rho_{\infty},\,x=0}
\equiv\frac{9}{2}\,\mathcal{E}^{(2,2)}\ ,
\label{ksym1}
\end{equation}
for the expansion (\ref{fmatter}). Here $K_{3}$ is the bulk anharmonicity
coefficient, $L_{\infty}$ is the density-symmetry coefficient (symmetry
energy slope parameter), $K_{\mathrm{sym}}$ is the symmetry energy curvature
parameter. Using (\ref{sigmaeq}), we write also 
\begin{equation}
%\[
\sigma\approx\sigma_{\infty}\left( 1-2\xi/R_{e}\right)\ ,\ \ \ \ \ 
%\]
%\begin{equation}
\sigma_{\infty}\approx\sigma _{0}+\sigma_{-}X^{2}\ ,
\ \ \ \xi \approx\xi_{0}+\xi_{-}X^{2}\ ,  \label{sigma1x}
\end{equation}
and, taking the advantage of the large mass limit
$E_{\infty}/A={E/A}\vert_{X=\mathrm{const},\,A\rightarrow\infty}$, one has
\begin{equation}
a_{V}=\mu _{\infty}\ ,
\ \ \ b_{V}=b_{\infty}\ .  \label{av1x}
\end{equation}
Using the conditions (\ref{eosmatt}) for the chemical potentials and both
relations (\ref{lambda01}) and (\ref{mu01p}), we obtain
%\[
$\rho/\rho_{\infty}\approx$ $1+6\,A^{-1/3}a_{S}/K_{\infty}-
3\,X^{2}L_{\infty}/K_{\infty}$
%X^{2}\left[ -\frac{3L_{\infty}}{K_{\infty }}+
%A^{-1/3}\left\{ \frac{6(b_{S}-2a_{S}L_{\infty}/K_{\infty})}{K_{\infty}}
%\left( 1-\frac{L_{\infty}}{b_{\infty }}\right)-\frac{6a_{S}}{K_{\infty}^{2}}
%\left[ L_{\infty}\left( 1-\frac{K_{3}}{K_{\infty}}\right)+
%K_{\mathrm{sym}}\right]\right\}\right]
%\] 
and 
\begin{equation}
a_{S}=4\pi r_{0}^{2}\sigma_{0}\ ,
\ \ \ b_{S}=4\pi r_{0}^{2}\left(\sigma_{-}+
\frac{2L_{\infty}}{K_{\infty}}\,\sigma_{0}\right)\ ,  
\ \ \ a_{c}=-8\pi r_{0}\sigma_{0}\left(\xi_{0}+
\frac{3\,\sigma_{0}}{K_{\infty}\rho_{\infty}}\right)\ ,
\label{as1x}
\end{equation}
%
%\begin{equation}
\[
b_{c}=-8\pi r_{0}\sigma_{0}\left\{\xi_{-}+\left(\frac{L_{\infty}}{K_{\infty}}+
\frac{\sigma_{-}}{\sigma_{0}}\right)\xi_{0}+
\frac{3\,\sigma_{0}}{K_{\infty}\rho_{\infty}}\left[\frac{L_{\infty}}{K_{\infty}}
\left( 4+\frac{K_{3}}{K_{\infty}}\right)-
\frac{K_{\mathrm{sym}}}{K_{\infty}}\right]+
\frac{3\,\sigma_{-}}{K_{\infty}\rho_{\infty}}\left( 2+
\frac{K_{\infty}\sigma_{-}}{2b_{\infty}\sigma_{0}}\right)\right\}\ .
%\label{bc1x}
%\end{equation}
\]
Here $r_{0}=(4\pi\rho_{\infty}/3)^{-1/3}$ and we have assumed $A^{-1/3}\ll 1$.
The equimolar, $R_{e}$, and Laplace, $R_{s}$,
radii defined by Eqs.~(\ref{evol}) and (\ref{p3}) read
\[
R_{e}\approx r_{0}A^{1/3}\left[ 1-
A^{-1/3}\frac{8\pi r_{0}^{2}\sigma_{0}}{K_{\infty}}
\right. 
\]
\begin{equation}
\left. 
+\ X^{2}\left[\frac{L_{\infty}}{K_{\infty}}-
A^{-1/3}\left\{\frac{8\pi r_{0}^{2}\sigma_{-}}{K_{\infty}}\left( 1-
\frac{L_{\infty}}{b_{\infty}}+\frac{K_{\infty }}{3\,\mu_{\infty}}\right)+
\frac{8\pi r_{0}^{2}\sigma_{0}}{K_{\infty}}\left[\frac{L_{\infty}}{K_{\infty}}\left( 3+
\frac{K_{3}}{K_{\infty}}\right) -\frac{K_{\mathrm{sym}}}{K_{\infty}}
\right]\right\}\right]\right]\ ,  \label{re1x}
\end{equation}
\[
R_{s}\approx r_{0}A^{1/3}\left[ 1-
A^{-1/3}\left(\frac{\xi_{0}}{r_{0}}+
\frac{8\pi r_{0}^{2}\sigma_{0}}{K_{\infty}}\right)
\right. 
\]
\begin{equation}
\left.
+\ X^{2}\left[\frac{L_{\infty}}{K_{\infty}}-
A^{-1/3}\left\{\frac{\xi_{-}}{r_{0}}+\frac{8\pi r_{0}^{2}\sigma_{-}}{K_{\infty}}\left( 1+
\frac{K_{\infty}}{2b_{\infty}}\frac{\sigma_{-}}{\sigma_{0}}\right)+
\frac{8\pi r_{0}^{2}\sigma_{0}}{K_{\infty}}\left[\frac{L_{\infty}}{K_{\infty}}\left( 3+
\frac{K_{3}}{K_{\infty}}\right)-\frac{K_{\mathrm{sym}}}{K_{\infty}}
\right]\right\}\right]\right]\ .  \label{rs1x}
\end{equation}
Using the derivations of $R_{e}$ and $R_{s}$, one obtains
\begin{equation}
R_{e}-R_{s}\approx\xi_{0}+\left[
\xi_{-}+\frac{3\sigma_{-}}{b_{\infty}\rho_{\infty}}\left(
\frac{\sigma_{-}}{\sigma_{0}}+\frac{2L_{\infty}}{K_{\infty}}-
\frac{2b_{\infty}}{3\mu_{\infty}}\right)\right] X^{2}=
\xi +\left[\frac{3\sigma_{-}}{b_{\infty}\rho_{\infty}}\left(
\frac{\sigma_{-}}{\sigma_{0}}+\frac{2L_{\infty}}{K_{\infty}}-
\frac{2b_{\infty}}{3\mu_{\infty}}\right)\right] X^{2}\ .
\label{R_e_s}
\end{equation}

To describe separately the neutron and proton density distributions we
introduce the neutron radius, $R_{n}$, and the proton radius, $R_{p}$, as
the dividing radii with zero value for the corresponding surface densities
$\varrho_{n,\mathcal{S}}=(\varrho_{\mathcal{S}}+\varrho_{-,\mathcal{S}})/2$ and
$\varrho_{p,\mathcal{S}}=(\varrho_{\mathcal{S}}-\varrho_{-,\mathcal{S}})/2$ :
\[
\left.\varrho_{n,\mathcal{S}}\right\vert_{R=R_{n}}=0\ ,\ \ \ \ \ \ 
\left.\varrho_{p,\mathcal{S}}\right\vert_{R=R_{p}}=0\ .
\]
The isovector shift of neutron-proton radii, $R_{n}-R_{p}$, is then written as
\begin{equation}
R_{n}-R_{p}\approx X\left[ -\frac{2\,\sigma_{-}}{b_{\infty}\rho_{\infty}}+
A^{-1/3}\left\{ 4\pi r_{0}^{2}\sigma_{0}\frac{4}{3b_{\infty}}\left(
\xi_{-}\!+\xi_{0}\frac{\sigma_{-}}{\sigma_{0}}\right)\! +4\pi r_{0}^{2}\sigma_{-}\!\left[
\frac{2\,\sigma_{-}}{b_{\infty}^{2}\rho_{\infty}}\!+\!\frac{4\,\sigma_{0}}
{b_{\infty}^{2}\rho_{\infty}}\left(\frac{L_{\infty}}{K_{\infty}}\!+\!
\frac{3b_{\infty}}{K_{\infty}}\right)\right]\right\}\right]\ .   \label{rnp1x}
\end{equation}
From Eq.~(\ref{rnp1x}) the value of neutron skin
$\sqrt{\langle r_n^2\rangle}-\sqrt{\langle r_p^2\rangle}$ is given within the
main order as
\begin{equation}
\sqrt{\langle r_n^2\rangle}-\sqrt{\langle r_p^2\rangle}\approx
-\sqrt{\frac{3}{5}}\,\frac{2\,\sigma_{-}X}{b_{\infty}\rho_{\infty}}=
\alpha X\ .
\label{skin}
\end{equation}
Here $\alpha=-2\sqrt{3/5}\,\sigma_{-}/(b_{\infty}\rho_{\infty})$
is the neutron skin parameter.
To describe the isospin dependence of surface energy within the droplet model
the effective surface stiffness, $Q$, have been introduced \cite{mysw69}.
At the large masses limit  $A\rightarrow\infty$ the droplet model result
reads
\begin{equation}
R_{n}-R_{p}\approx\frac{3}{2}r_{0}\frac{b_{\infty}}{Q}\,X\ .\label{dropletrnp}
\end{equation}
Using the main term on the right side of Eq.~(\ref{rnp1x}) together with Eq.~(\ref{dropletrnp})
one obtains the surface stiffness $Q$ as
\begin{equation}
Q=-\frac{9\,b_{\infty}^2}{16\pi r_{0}^{2}\,\sigma_{-}}\ .
\label{dropletQ}
\end{equation}
The values of $\alpha$ and $Q$ for different Skyrme forces are given in
Table~\ref{tab3}.

\noindent
\begin{table}
\begin{minipage}{0.53\textwidth}
\caption{Nuclear bulk parameters for different Skyrme forces.
The planar surface values $\sigma_{0}$, $\sigma_{-}$ and
$\xi_{0}$, $\xi_{-}$ were obtained by extrapolation $A\rightarrow\infty$,
see Fig.~\ref{fig5}.\label{tab1}}
\begin{tabular}{lllll}
\hline\hline\noalign{\smallskip}
& ~SkM~~ & ~SkM*~ & SLy230b$\!\!\!\!\!$ & ~~~T6 \\
\noalign{\smallskip}\hline\noalign{\smallskip}
$\mu_{\infty}$ (MeV)                & -15.77 & -15.77 & -15.97 & -15.96 \\
$\rho_{\infty}$ (fm$^{-3}$)        & ~0.1603 & ~0.1603 & ~0.1595 & ~0.1609 \\
$K_{\infty}$ (MeV)                 & ~216.6  & ~216.6  & ~229.9  & ~235.9 \\
$K_{3}$ (MeV)                         & ~913.5 & ~913.5 & ~1016. & ~1032. \\
$K_{\mathrm{sym}}$ (MeV)           & -148.8 & -155.9 & -119.7 & -211.5 \\ 
$b_{\infty}$ (MeV)                   & ~30.75 & ~30.03 & ~32.01 & ~29.97 \\
$L_{\infty}$ (MeV)                  & ~49.34 & ~45.78 & ~45.97 & ~30.86 \\
$\sigma_{0}$ (MeV/fm$^{2}$) & ~0.9176 & ~0.9601 & ~1.006 & ~1.021 \\
$\xi_{0}$ (fm)                       & -0.3565 & -0.3703 & -0.3677 & -0.3593 \\
$\sigma_{-}$ (MeV/fm$^{2}$)$\!\!\!\!$  & -3.118 & -3.094 & -3.131 & -2.413 \\
$\xi_{-}$ (fm)                      & -5.373 & -5.163 & -4.590 & -2.944 \\
\noalign{\smallskip}\hline
\end{tabular}
\end{minipage}
\hfill
\begin{minipage}{0.46\textwidth}
\caption{Mass formula coefficients for finite nuclei.\label{tab2}}
\begin{tabular}{lllll}
\hline\hline\noalign{\smallskip}
& ~SkM & ~SkM*$\!\!$ & SLy230b$\!\!\!\!\!\!$ & ~~T6 \\
\noalign{\smallskip}\hline\noalign{\smallskip}
$a_{V}$ (MeV) & -15.8 & -15.8 & -16.0 & $\!\!$-16.0 \\
$a_{S}$ (MeV) & ~15.0 & ~15.7 & ~16.5 & $\!\!$~16.7 \\
$a_{c}$ (MeV) & ~7.30 & ~7.92 & ~8.26 & $\!\!$~8.16 \\
$b_{V}$ (MeV) & ~30.8 & ~30.0 & ~32.0 & $\!\!$~30.0 \\
$b_{S}$ (MeV) & -44.2 & -44.1 & -44.9 & $\!\!$-35.1 \\
$b_{c}$ (MeV) & ~35.7 & ~35.1 & ~28.6 & $\!\!$~17.3 \\
$r_{S/V}\!=\!|b_{S}/b_{V}|\!\!\!$ & ~1.44 & ~1.47 & ~1.40 & $\!\!$~1.17 \\
\noalign{\smallskip}\hline
\end{tabular}\vspace*{3ex}
\caption{Neutron skin parameter $\alpha$ and surface stiffness $Q$ for
different Skyrme forces. The values of $\alpha$ and $Q$ were calculated
using Eqs.~(\ref{skin}) and (\ref{dropletQ}), respectively.\label{tab3}}
\begin{tabular}{lllll}
\hline\hline\noalign{\smallskip}
& ~SkM & ~SkM* & SLy230b$\!\!\!\!$ & ~~T6 \\
\noalign{\smallskip}\hline\noalign{\smallskip}
$\alpha$ (fm) & ~0.980 & ~0.996~ & ~0.950~ & ~0.775~ \\
$Q$ (MeV) & ~~41.6 & ~~40.0~ & ~~44.8~ & ~~51.2~ \\
\noalign{\smallskip}\hline
\end{tabular}
\end{minipage}
\end{table}

\section{CONCLUSIONS}
We propose a new method of the evaluation of the $A$-dependency of
$\beta$-stability line and both the Coulomb, $e_{C}(A)$, and the symmetry,
$b(A)$, energies. Our method is model independent in a sense that
it does not imply a theoretical model for the calculation of the nuclear
binding energy. The method is based on the experimental data for the shift
of the neutron-proton chemical potential $\Delta\lambda (A,X)$ for nuclei
beyond $\beta$-stability line but at the fixed total particle number $A$.
We show the presence of the thin structure (sawtooth shape) of $\beta$-stability
line for the curve $X^{\ast}(A)$ which is not observed at the
traditional presentation of $\beta$-stability line as $Z(N)$-dependency. We
note that this non-monotonic behavior of $\beta$-stability line is the
consequence of subshell structure of single particle levels near Fermi
energy for both the neutrons and the protons.
%Due to the subshell structure,
%the Fermi energies for protons and neutrons can coincide by chance only
%creating the non-monotonic behavior of $X^{\ast}(A)$.
We demonstrate the correlation between the positions of maxima of function
$X^{\ast}(A)$ and double magic nucleon numbers.

We have suggested the model independent method for calculation of
the Coulomb energy parameter $e_{C}(A)$ which absorbs both the finite
diffuse layer and the quantum exchange contributions. The last fact leads to
more complicate $A$-dependence of $e_{C}(A)$ (see Eq.~(\ref{ec})) than the
traditional one $e_{C}(A)\propto A^{2/3}$. We have established the
dependence of $\beta$-stability line $X^{\ast}(A)$ on the Coulomb,
$e_{C}(A)$, and the symmetry, $b(A)$, energies. That allowed
us to redefine a smooth $A$-dependency of $\beta$-stability line
(see Eq.~(\ref{xx})) which can be used instead the phenomenological one (\ref{xstar})
given by Green and Engler \cite{gren53}.
One should note that it is difficult to determine $b_{V}$
unambiguously. The reason is that the surface contribution cannot be neglected
even for the heavy nuclei covered by the experimental data. Another reason is
the different $A$-dependence for $b(A)$ used in the nuclear mass
formula, see e.g. Ref.~\cite{dani03}.

We have also observed the thin structure of the symmetry energy
$b(A)$. The value of $b(A)$ has the canyon-like
$A$-dependence for a fixed neutron excess $A_{-}=N-Z$. The width and the position
of bottom for such a canyon-like shape depend on the neutron excess and are
related to the subshell structure in the discrete spectrum of the single
particle levels for both the neutrons and the protons. The canyon shape of
$b(A)$ becomes thinner and deeper near double (proton-neutron)
magic number.

Considering a small two-component, charged droplet with a finite diffuse
layer, we have introduced a formal dividing surface of radius $R$ which
splits the droplet onto volume and surface parts. The corresponding
splitting was also done for the binding energy $E$. Assuming that the
dividing surface is located close to the interface, we are then able to
derive the surface energy $E_{S}$. In general, the surface energy
$E_{S}$ includes the contributions from the surface tension $\sigma$
and from the binding energy of $A_{S}$ particles located within
the surface layer. The equimolar surface and thereby the actual physical
size of the droplet are derived by the condition
$\varrho_{S}\lambda +\varrho_{-,S}\lambda_{-}=0$
which means that the latter contribution is excluded from the surface energy providing
$E_{S}\propto\sigma$.

In a small nucleus, the diffuse layer and the curved interface affect the
surface properties significantly. In agreement with Gibbs--Tolman concept 
\cite{tolm49,gibbs}, two different radii have to be introduced in this case.
The first radius, $R_{s}$, is the surface tension radius (Laplace radius)
which provides the minimum of the surface tension coefficient $\sigma$ and
the fulfillment of the Laplace relation (\ref{p3}) for capillary pressure.
The another one, $R_{e}$, is the equimolar radius which corresponds to the
equimolar dividing surface due to the condition (\ref{emolar}) and defines
the physical size of the sharp surface droplet, i.e., the surface at which
the surface tension is applied. The difference of two radii $R_{e}-R_{s}$ in
an asymptotic limit of large system $A\rightarrow\infty$ derives the
Tolman length $\xi$. That means the presence of curved surface is not
sufficient for the presence of the curvature correction in the surface
tension. The finite diffuse layer in the particle distribution is also
required. We point out that the Gibbs--Tolman theory allows to treat a liquid
drop within thermodynamics with minimum assumptions. Once the binding energy
and chemical potential of the nucleus are known its equimolar radius, radius
of tension and surface energy can be evaluated using the equation of state
for the infinite nuclear matter.
%For a symmetric liquid the
%value of Tolman length is about of half of the diffuseness parameter $a$ for
%the nuclear surface layer.
We have also established the relation of the
macroscopic energy coefficients in the liquid drop model expansion
Eq.~(\ref{eldm1}) to the nuclear matter parameters.

The sign and the magnitude of the Tolman length $\xi$ depend on the
interparticle interaction. We have shown that the Tolman length is negative
for a nuclear Fermi liquid drop. As a consequence, the curvature correction
to the surface tension leads to the hindrance of the yield of light
fragments at the nuclear multifragmentation in heavy ion collisions. We have
also shown that the Tolman length is sensitive to the neutron excess and its
absolute value growth significantly with growing asymmetry parameter $X$.

\end{document}